\documentclass{ws-mpla}
\usepackage[super]{cite}
\usepackage[utf8]{inputenc}
\usepackage{bm}
\usepackage{amssymb}
\usepackage{amsbsy}
\usepackage{physics}
\usepackage{xcolor}
\begin{document}
\markboth{Rasouli, Jalalzadeh, Moniz}
{Broadening  quantum cosmology with a fractional whirl}
\newcommand{\lyxline}[1][1pt]{%
  \par\noindent%
  \rule[.5ex]{\linewidth}{#1}\par}

\makeatother
\newcommand{\bl}{\textcolor{blue}}
\newcommand{\red}{\textcolor{red}}

\catchline{}{}{}{}{}
\title{Broadening quantum cosmology with a fractional whirl
}

\author{S. M. M. Rasouli}
\address{Departamento de F\'{i}sica,\\
Centro de Matem\'{a}tica e Aplica\c{c}\~{o}es (CMA - UBI), \\
Universidade da Beira Interior,\\
 Rua Marqu\^{e}s d'Avila
e Bolama, 6200-001 Covilh\~{a}, Portugal. \\ and \\ 
Department of Physics, \\ Qazvin Branch, \\ Islamic Azad University, \\ Qazvin, Iran.
\\ mrasouli@ubi.pt}

\author{S. Jalalzadeh}
\address{Departmento de Física, \\ Universidade Federal de Pernambuco, \\ Pernambuco, PE 52171-900, Brazil\\ shahram.jalalzadeh@ufpe.br}
\date{November 2020}

\author{P. V. Moniz}
\address{Departamento de F\'{i}sica,\\
Centro de Matem\'{a}tica e Aplica\c{c}\~{o}es (CMA - UBI), \\
Universidade da Beira Interior,\\
 Rua Marqu\^{e}s d'Avila
e Bolama, 6200-001 Covilh\~{a}, Portugal. \\
pmoniz@ubi.pt
}


\maketitle
\pub{Received (Day Month Year)}{Revised (Day Month Year)}

\begin{abstract}
We start by presenting a brief summary of fractional quantum mechanics, as
means to convey a motivation towards {\it fractional quantum cosmology}. Subsequently, such application
is made concrete with the assistance of a case study. Specifically, we investigate and then discuss a  model of stiff matter in a spatially flat homogeneous and isotropic universe. A {\it new} quantum cosmological solution, where fractional calculus implications are 
explicit, is presented and then contrasted with 
the corresponding standard quantum cosmology setting. 

\end{abstract}

\ccode{PACS Nos.:  05.40.Fb, 04.60.-m, 98.80.Qc}

\section{Introduction}
\label{Int}

\indent

A particular branch of mathematical analysis, albeit not yet widely explored,  is  fractional calculus,  in which the power of the differentiation operator can be any rational  (or even  a real or a complex) number. Indeed, investigations into fractional calculus \cite{OS74,SKM99,Herrmann,KST06,P99,A} started in the 17th century, which is almost the same temporal frame as the integer-order calculus started to develop. However, in the early stages of the evolution of these two branches of mathematics, due to the lack of practical evidence in physics (and mechanics, specifically), the former developed much more slowly
 than the latter. This was the setting until Mandelbrot \cite{M82} realized that
an exciting  connection between fractional Brownian motion and Riemann-Liouville fractional calculus could be reasonably established. After that remark, applications of  fractional calculus to study phenomena in engineering as well as other various fields of science have increased, and therefore the equations of fractional calculus  have played a significant role in describing, e.g., anomalous transport, super-slow relaxation, diffusion-reaction processes\cite{MK00, MLP01, Z02, Z05, CM97}. In addition, areas of contemporary engineering were also explored within fractional
calculus\cite{Ortigueira-1}.

Furthermore, the physical world includes not only the conservative
systems but also the non-conservative ones (due to the presence of
friction). Advanced procedures like  classical
Hamiltonian (or Lagrangian)
mechanics (which is formulated based on the derivative of integer order) could be replaced by fractional Hamiltonian (and Lagrangian) equations of motion,
aiming in
investigating analytically any
peculiar consequences of the  non-conservative forces \cite{S06,R96,R97,A02,A06,BA06,BMT06,RAMB08,MBR06,C67,T08}.

Similarly, in the context of quantum physics, there might be suggestive physical reasons to reformulate the theory based on  fractional calculus. For instance, if restricting to only consider the Brownian paths, it is impossible to analyze other pertinent many 
phenomena\cite{L00}. Such consideration in quantum physics leads to generalized versions of the well-known Schr\"{o}dinger equation (SE). More concretely, by including non-Brownian trajectories
in the path integral derivation, either the {\it space}-fractional, {\it time}-factional and {\it space-time}-factional versions of the standard SE have been established \cite{L00,L00-0,L00-1,L02,L17,B,N04,WX07,DX08,I09,W10,I16,GX06,DX07,T1,T2}. Let us introduce them briefly and then,  in the next section, we further specify (though shortly) about their formalism:

\begin{itemize}
  \item
   In deriving the space-fractional SE, the Feynman-Hibbs path integral procedure \cite{FH65} has been extended by Laskin such that the Gaussian probability distribution was replaced by L\'{e}vy's\cite{L00, L00-0, L00-1, L02}.
More precisely, the second-order spatial derivative in the standard SE is then
modified to the fractional-order $\alpha$, whilst the time sector remains unchanged. This equation is characterized by the L\'{e}vy index
$\alpha$, $0<\alpha\leq2$, such that for the particular case where $\alpha=2$, the standard SE is recovered.
Moreover, the Laplacian operator acquires a fractional Riesz derivative\cite{R49}.

By applying some examples of  specific potential fields, this (space) fractional quantum system has been coherently exlored\cite{GX06, DX07}.

\item
From employing the Caputo fractional derivative 
\cite{P99, Ortigueira-1, KST06}, the first-order time derivative associated with the standard SE has been generalized to being
fractional. Therefore, the time-fractional SE was obtained \cite{N04}. The associated Hamiltonian  is non-Hermitian and not local in time. Indeed, in this procedure, analogous to the map between the standard versions of the diffusion equation and the SE, the time-fractional diffusion equation was also mapped into the time-fractional SE \cite{L17}. We should note that, in the latter, the spatial derivative, the same as the standard SE, is still the second order. Indeed, unlike the standard SE and the space-fractional SE (which both obey the Markovian evolution law), the time-fractional SE describes a non-Markovian evolution in quantum physics \cite{DX08}. As applications of the time-fractional SE, a free particle and a particle in a particular potential well have been investigated \cite{N04}.

 \item
 The term ``space-time fractional SE''  has been used for the first time by Wang and
 Xu\cite{WX07}, where they employed a combination of the procedures used by
 Naber\cite{L00} and
 Laskin\cite{N04} to establish this equation. More precisely, this is the most generalized version of the standard SE, which is retrieved by including both the Caputo fractional derivative and the quantum Riesz fractional operator\cite{R49}.  The space-time fractional QM was applied to study a free particle and an infinite square potential well \cite{WX07}.  Moreover,  Dong and Xu\cite{DX08} have represented the work of Wang and Xu (although with a minor difference) and then applied the space-time fractional SE to study a quantum particle in a $\delta$-potential well.  Furthermore,   not only the space-time fractional SE has recently been represented by Laskin\cite{L17} within  his own method but also the important features of the fractional QM approaches have been outlined.
 In the therein established framework\cite{L17}, the space-time
 fractional SE includes not only dimensional parameters but also a set  of dimensionless fractality parameters. These latter  parameters are responsible for modeling the spatial and temporal fractality \cite{L17}.
\end{itemize}

Recently, fractional quantum mechanics (FQM)
has been considered as a tool to explore features
within cosmology. Namely, {\it fractional quantum cosmology} (FQC). It has been
emerging gradually\cite{GC,JF,mdpi}, subsequently  pointing to fascinating
opportunities and `lateral' connections with what were unforeseen
mathematics and physics domains, with  this review herewith following a
particular (canonical) route as charted in another publication\cite{mdpi}.

Before proceeding, let us point that this review
paper is organized as follows. In Section \ref{FQM}, we briefly
review  fractional quantum mechanics (publications with much more detail can be found in the bibliography\cite{L17,B,L00-0,L00}). Subsections \ref{S-FQM} and \ref{ST-FQM}
convey the summary line of FQM
extensions within {\it spatial}-FQM and {\it space-time}-FQM, respectively.  Section \ref{FQC}
reports the  study of a cosmological model within the 
 scope of {\it time}-FQM. Finally, in section \ref{DandO} we
present a brief discussion and outlook with respect canonical
FQC,
as unveiled in this review paper and a  recent contribution as well\cite{mdpi}.

\section{ A brief review of  fractional quantum mechanics}
\label{FQM}

\indent

In this section, we would like to introduce and summarize  the space-fractional SE and the space-time fractional SE. In order to retrieve  more details of the mentioned frameworks and their applications, we suggest the readers consider
additional references, indicated in our bibliography\cite{L00,L00-0,L00-1,L02,N04,WX07,DX08}.

\subsection{Space fractional quantum mechanics}
\label{S-FQM}

\indent

The Hamiltonian of a system in  classical mechanics is given by
\begin{equation}
    H(\mathbf{p}, \mathbf{r}) :=
\frac{\mathbf{p}^2}{2m}
+ V (\mathbf{r}),
\label{2}
\end{equation}
where $\mathbf{r}$ and $\mathbf{p}$, respectively, stand for
the space coordinates and the corresponding
momentum,  associated with a particle with mass $m$;
the potential energy, in general, is assumed to
be a function of the space coordinates, i.e., $V=V (\mathbf{r})$.
In analogy with  classical mechanics, the Hamiltonian of
 quantum mechanics can  be extracted from  \eqref{2}, provided
that the corresponding quantities are
taken as operators (which are denoted with an over-hat):
\begin{equation}
   \hat{H}(\hat{\mathbf{p}}, \hat{\mathbf{r}}) :=
\frac{\hat{\mathbf{p}}^2}{2m}
+ \hat{V} (\hat{\mathbf{r}}).
\label{2a}
\end{equation}

Pondering the empirical physical (classical) fact realized in the square dependence
of the momentum in equations \eqref{2} and \eqref{2a}, such feature  has
motivated researchers to explore diverse functions of
the kinematic term, such that it would be nevertheless consistent with the fundamental
principles of classical mechanics and quantum mechanics \cite{B}.
Such an investigation was launched for the first time by Laskin, who
chose the Feynman path integral approach (to quantum mechanics) as an appropriate procedure.
Let us be more precise. In the Laskin approach, the Brownian-like quantum
mechanical trajectories applied in the
Feynman's framework, are replaced by L\'{e}vy-like ones.
\footnote{Although, it should be noted that the path integral
 over the L\'{e}vy paths has formerly been brought up by Kac\cite{K51}. }
More concretely, applying a natural generalization\cite{L00,L02,B}
of equation \eqref{2} as
\begin{equation}
    H_\alpha
(\mathbf{p}, \mathbf{r}) : = D_\alpha
|\mathbf{p}|^\alpha
+ V (\mathbf{r}), \hspace{10mm} 1 < \alpha \leq  2,
\label{5}
\end{equation}
(where $D_\alpha$ is taken as a generalized coefficient carrying
dimension $[D_\alpha]={\rm erg}^{1-\alpha}{\rm cm}^\alpha{\rm sec}^{-\alpha}$)
and letting ${\mathbf{p}}\rightarrow \hat{\mathbf{p}}$
and ${\mathbf{r}}\rightarrow \hat{\mathbf{r}}$, the (space) fractional quantum dynamics is constructed.
Obviously, in the particular case where $\alpha  = 2$ and $D_\alpha = 1/(2m)$,
equation \eqref{5} reduces to \eqref{2}.
Namely, the fractional quantum mechanics (FQM) established
through applying the L\'{e}vy path integral is a natural
generalization of the Feynman path integral.
Consequently, the space-fractional SE (including the space derivatives of order $\alpha$),
which is a generalized version of the standard one, is obtained via
applying the Feynman path integral over Brownian
trajectories\cite{B}.
In summary, in the particular case where $\alpha=2$, the equations
of the standard (non-fractional) quantum mechanics are
recovered from the corresponding ones associated
with the space-fractional quantum mechanics.

Now, we briefly introduce the space-fractional SE.
In analogy with the same procedure that yields the standard SE, let us admit
\begin{equation}
  E:=
D_\alpha
|\mathbf{p}|^\alpha
+ V (\mathbf{r}), \hspace{10mm} 1 < \alpha \leq  2,
\label{E}
\end{equation}
where $E$ and $\mathbf{p}$ can transform as
\begin{equation}
    E \rightarrow  i\hbar \frac{\partial }{\partial  t}, \hspace{10mm}
    \mathbf{p} \rightarrow  -i\hbar \mathbf{\nabla},
    \label{6}
\end{equation}
where  $\nabla = \frac{\partial }{\partial  \mathbf{r}}$
and $\hbar$ is the Planck' s constant over $2\pi$.
Therefore, the space-fractional SE can be obtained by
employing the well-known transformations in \eqref{6} into equation \eqref{5} and
using them\cite{L00,L02,B}
to eventually extract the wave function $\psi(\mathbf{r},t)$ :
\begin{equation}
    i\hbar \frac{\partial \psi(\mathbf{r},t)}{\partial  t}=
D_\alpha
(-\hbar^2 \Delta)^{\alpha/2} \psi(\mathbf{r},t)
+ V(\mathbf{r},t) \psi(\mathbf{r},t), \hspace{10mm} 1 < \alpha \leq  2.
\label{7-0}
\end{equation}
In equation \eqref{7-0}, $\psi(\mathbf{r},t)$ and $\triangle := \nabla.\nabla$ are the wave function in space representation and the Laplacian, respectively. Moreover, the
 fractional (quantum) Riesz derivative\cite{R49,B} in $3D$, $(-\hbar^2 \Delta)^{\alpha/2}$, is given by
 \begin{equation}
(-\hbar^2 \Delta)^{\alpha/2} \psi(\mathbf{r},t)
=
\frac{1}{(2\pi\hbar)^3}
\int d^3 p e^{i\frac{\mathbf{p}\cdot \mathbf{r}}{\hbar}}
|\mathbf{p}|^\alpha
\varphi(\mathbf{p},t),
\label{8}
\end{equation}
where $\varphi(\mathbf{p},t)$ is the wave function in
momentum representation, which is related to $\psi(\mathbf{r},t)$ by $3D$ Fourier transforms.

In summary, replacing ${\mathbf{r}}$ and ${\mathbf{p}}$ by quantum
mechanical operators $\hat{\mathbf{r}}$ and $\hat{\mathbf{p}}$, equation \eqref{5} leads
to the fractional Hamilton operator $\hat{H}_\alpha(\hat{\mathbf{p}}, \hat{\mathbf{r}})$:
\begin{equation}
    \hat{H}_\alpha
(\mathbf{\hat{p}}, \mathbf{\hat{r}}) : = D_\alpha
|\mathbf{\hat{p}}|^\alpha
+ V (\mathbf{\hat{r}}), \hspace{10mm}  1 < \alpha \leq  2.
\label{5-1}
\end{equation}
Therefore, the space-fractional SE
equation can also be rewritten in the operator form as
\begin{equation}
    i\hbar \frac{\partial \psi(\mathbf{r},t)}{\partial  t}
= \hat{H}_\alpha(\hat{\mathbf{p}}, \hat{\mathbf{r}}) \psi(\mathbf{r},t).
\label{7}
\end{equation}

In what follows, it is worthy to present some features
of   FQM \cite{B}, which are considered as the generalized version
of those known in the standard quantum mechanics:
(i) The fractional Hamiltonian operator is a self-adjoint or
Hermitian operator in the space with scalar product \cite{B};
 (ii) The parity conservation law for the FQM is given
 by $\mathbf{\hat{P}}H_\alpha=H_\alpha \mathbf{\hat{P}}$,
 where $\mathbf{\hat{P}}$ denotes the inversion operator;
 (iii) It has been shown that
 \begin{equation}
    \frac{\partial \rho(\mathbf{r},t)}{\partial  t}+{\rm div} \mathbf{j}(\mathbf{r},t)=0,
\label{ro-j}
\end{equation}
 where $\rho(\mathbf{r},t):=  \overset{*}{\psi}(\mathbf{r},t)\psi(\mathbf{r},t)$
 and $\mathbf{j}(\mathbf{r},t)$ denote the quantum mechanical probability density
 and the fractional probability current
 density vector, respectively, where the latter is defined as
 \begin{equation}
     \mathbf{j}:= \frac{1}{\alpha}\left(\psi \mathbf{\hat{v}}
     \overset{*}{\psi}+\overset{*}{\psi}\mathbf{\hat{v}}\psi\right),\hspace{10mm}1 < \alpha \leq  2.
\label{j}
\end{equation}
Notwithstanding the formal similarities with the
standard formulation, in equation \eqref{j} the velocity operator $\mathbf{\hat{v}}=d\mathbf{\hat{r}}/dt$ is specifically given instead by
 \begin{equation}
    \mathbf{\hat{v}}=\alpha D_\alpha |\mathbf{\hat{p}^2}|^{\alpha/2-1}\mathbf{\hat{p}}.
\label{v}
\end{equation}

Let us close this subsection by focusing on the special case where the
 Hamiltonian does not depend explicitly on time. In this case,
 the space-fractional SE \eqref{7-0}  is satisfied by the special solution and in one dimension
\begin{equation}
    \psi(x, t) = \left[ \exp\left( - \frac{iEt}{\hbar}
    \right)\right]
\phi(x), \label{8a}
\end{equation}
provided that
\begin{equation}
    H_\alpha \phi(x) = -D_\alpha(\hbar \nabla)^\alpha \phi(x) + V(x) \phi(x)= E \phi(x),
     \hspace{10mm} 1 < \alpha \leq 2,
     \label{9}
\end{equation}
which is the time-independent fractional SE \cite{B}.
Equation \eqref{9} implies that the probability
 to find the particle at $x$ is equal to $|\phi|^2$, which is independent of time.

 \subsection{Space-time fractional quantum mechanics}
\label{ST-FQM}

\indent

As mentioned,  the space-time fractional SE has been
originally established by Wang and
Xu \cite{WX07} and then further discussed  by Dong and Xu \cite{DX08}.
In this subsection, let us introduce  this interesting framework
from within the approach of Laskin \cite{L17}, in which the wording
``time fractional QM'' has been used because the time derivative of the SE associated
with both the ``standard QM'' and the ``space FQM'' is substituted by a
fractional time derivative (namely, the Caputo fractional time derivative).

The space-time fractional SE can be
considered as a generalized version of equation \eqref{7}:
\begin{equation}
   \hbar_\beta i^\beta \partial_t^\beta \psi(\mathbf{r},t)
= \hat{H}_{\alpha,\beta}\left(\hat{\mathbf{p}}_\beta, \hat{\mathbf{r}}\right)
\psi(\mathbf{r},t),   \hspace{10mm} 1<\alpha\leq2, \hspace{5mm}0<\beta \leq1,
\label{ST-1}
\end{equation}
where
\begin{equation}
\hat{H}_{\alpha,\beta}\left(\hat{\mathbf{p}}_\beta, \hat{\mathbf{r}}\right)
\: = D_{\alpha,\beta}
|\mathbf{\hat{p}}_\beta|^\alpha
+ V (\mathbf{\hat{r}},t),
\label{ST-2}
\end{equation}
in which, $ \hat{\mathbf{r}}$ and $\hat{\mathbf{p}}=-i\hbar_\beta\frac{\partial}{\partial \mathbf{r}}$
denote the $3D$ quantum operator of coordinate and $3D$ time fractional quantum momentum operator, respectively.
Moreover, in \eqref{ST-1}, $ \hbar_\beta$ and $D_{\alpha,\beta}$ are
 two scale coefficients with physical dimensions $[\hbar_\beta]={\rm erg}.{\rm sec}^\beta$ and
  $[D_{\alpha,\beta}]={\rm erg}^{1-\alpha}.{\rm cm}^\alpha.{\rm sec}^{-\alpha\beta}$; $\partial_t^\beta$
denotes the left Caputo fractional derivative \cite{C67} of order $\beta$:
\begin{equation}
 \partial_t^\beta f(t)=\frac{1}{\Gamma(1-\beta)}\int_{0}^{t}d\tau\frac{f'(\tau)}{(t-\tau)^\beta},\hspace{10mm} 0<\beta \leq1,
\label{ST-3}
\end{equation}
where $f'(\tau)\equiv \frac{df(\tau)}{d\tau}$.
It should be reminded that in the time fractional QM, the operator $\hat{H}_{\alpha,\beta}\left(\hat{\mathbf{p}}_\beta, \hat{\mathbf{r}}\right)$
introduced by equation \eqref{ST-2} is not the Hamilton operator of the
considered quantum mechanical system, and it was therefore called the
pseudo-Hamilton operator \cite{DX08}. However, it is straightforward to show that $\hat{\mathbf{r}}$ and $\hat{\mathbf{p}}_\beta$
are Hermitian operators \cite{L17}.
Moreover, assuming $V (\mathbf{\hat{r}},t)=V (-\mathbf{\hat{r}},t)$, we find that the parity conservation law is satisfied,
i.e., $\mathbf{\hat{P}}\hat{H}_{\alpha,\beta}=\hat{H}_{\alpha,\beta}\mathbf{\hat{P}}$.

In this review, we abstain from detailing  the applications of the three frameworks of FQM, since much is available in the literature and we point to some in the
bibliography. However, in what follows, let us mention a few general features\cite{L17}:

\begin{enumerate}
  \item
  In special cases, the most generalized FQM yields the following well-known cases (let us focus on a one-dimensional case):
   \begin{itemize}
  \item
  For $\alpha=2$ and $\beta=1$, we have $D_{\alpha,\beta}=1/(2m)$, $p_\beta=p$ and $\hbar_\beta=\hbar$,
  therefore the traditional SE is recovered.

  \item
  Assuming $1<\alpha\leq2$ and $\beta=1$, we find that
   $D_{\alpha,\beta}\rightarrow D_{\alpha}$, and
   therefore the space fractional SE \cite{L02} is recovered.

   \item
   Assuming $\alpha=2$ and $0<\beta \leq1$, we find that $D_{\alpha,\beta}\rightarrow D_{2,\beta}$
   and therefore the time fractional SE is retrieved,   which can be considered as an alternative route  to that established by Naber \cite{N04}.
  \end{itemize}

\item
Notwithstanding the previous item,  let us outline a few shortcomings of the time fractional QM.
It has been shown that the space fractional QM \cite{L00,L02,B} supports all the QM fundamentals. Whilst, the following QM
characteristics are violated by the time fractional QM:
(i) Quantum superposition law; (ii) Unitarity of
evolution operator; (iii) Probability conservation law;
(iv) Existence of stationary energy levels of quantum system.
Moreover, as mentioned, in spite of the standard and
the space fractional QM, whose dynamics are governed by a
 Hamilton operator, the operator $\hat{H}_{\alpha,\beta}$
associated with the time fractional QM is a pseudo-Hamilton (namely, its eigenvalues are not the
energy levels of a time fractional quantum system).

\item
Let us here remark a few advantages of the time fractional QM.
Indeed, investigating it  highlighted the importance and fundamental beauty of the   standard as well as space fractional QM (some of them mentioned above); the time fractional QM is however a great challenge for studying and discovering new mathematical tools,  which have not yet been applied in the the standard as well as space fractional QM. Concretely, in order to investigate dissipative quantum systems interacting with environment, the time fractional QM can be considered as an enticing  approach \cite{T08,I09,W10,I16}.
\end{enumerate}

\section{ Fractional quantum cosmology case study}
\label{FQC}

\indent

As a simple model of application of FQM towards canonical quantum cosmology\cite{C2,VargasMoniz:2020hve,moniz2010quantum-a,moniz2010quantum-b} (i.e.,
(canonical) fractional quantum cosmology\cite{mdpi}), let us consider a Friedmann-Lema\^itre-Robertson-Walker (FLRW) universe
with a line element
\begin{eqnarray}\label{sh1}
ds^2=-N^2(t)dt^2+e^{2x(t)}\left[\frac{dr^2}{1-kr^2}+r^2d\Omega^2\right],
\end{eqnarray}
where $N(t)$ is the lapse function, $e^{x(t)}$ represents the scale factor and $k = \pm1, 0$ is the spatial 3-curvature of a homogeneous and isotropic 3-dimensional compact and without boundary hypersurface, $\Sigma_t$.
Note that the compactness of $\Sigma_t$ requires that the 3-volume $V_k$ is finite. The ADM action functional of the
gravitational sector plus matter fields (in this work, we will consider a perfect fluid with the barotropic equation of state $p=\omega \rho$, where $\omega$ is a constant and $\rho$, $p$ stand for the energy density and pressure, respectively) in the formalism developed by Schutz \cite{Schutz} is
\begin{eqnarray}\label{sh2}
\begin{array}{cc}
S=\frac{1}{16\pi G}\displaystyle\int_{t_i}^{t_f}dt\int_{\Sigma_t}d^3xN\sqrt{h}\left[~^{(3)}R+K_{ij}K^{ij}-K^2\right]\\+\displaystyle\int_{t_i}^{t_f}dt\int_{\Sigma_t}d^3xN\sqrt{h}p,
\end{array}
\end{eqnarray}
where $^{(3)}R$, $K_{ij}$ and $h_{ij}$ denote the Ricci scalar, the extrinsic curvature, and the induced metric of $\Sigma_t$, respectively. In Schutz's formalism, the fluid’s 4-velocity can be expressed in terms of five potentials $\sigma, \zeta, \bar\beta,\theta$ and $\mathcal S$
\begin{eqnarray}\label{sh3}
u_\nu:=\frac{1}{\mu}(\epsilon_{,\nu}+\zeta{\bar\beta}_{,\nu}+\theta{\mathcal S}_{,\nu}),~~~~~u_\nu u^\nu=-1,
\end{eqnarray}
where $\mu$ denotes the specific enthalpy of the fluid, $\mathcal S$ is the specific entropy, and the potentials $\zeta$ and $\bar\beta$ are connected with
rotation which are absent of homogeneous and isotropic models.
The action (\ref{sh2}) corresponding to the line element (\ref{sh1}), after some thermodynamical considerations and using the constraints for the fluid\cite{Rubakov}, reduces to
\begin{eqnarray}\label{sh4}
\begin{array}{cc}
S=\displaystyle\int_{t_i}^{t_f}\left[\frac{3\mathcal V_k}{8\pi G}\left(-\frac{e^{3x}\dot x^2}{N}+kNe^x\right)+\frac{\omega\mathcal V_k e^{3x}e^{-\frac{\mathcal S}{\omega}}}{(1+\omega)^{1+\frac{1}{\omega}}N^\frac{1}{\omega}}(\dot\epsilon+\theta\dot{\mathcal S})^{1+\frac{1}{\omega}}\right]dt,
\end{array}
\end{eqnarray}
where an overdot denotes differentiation with respect time coordinate $t$. It is straightforward to show that the ADM Hamiltonian of our herein model is given by
\begin{eqnarray}\label{sh5}
\begin{array}{cc}
H_\text{ADM}=\dot x p+\dot{\mathcal S}p_{\mathcal S}+\dot\epsilon p_\epsilon-L_\text{ADM}\\
\\
=\tilde N\left[-\frac{2\pi}{m_\text{P}^2\mathcal V_k}e^{3(\omega-1)x}p^2+\frac{e^{\mathcal S}}{{\mathcal V_k}^\omega}p_\epsilon^{1+\omega}-\frac{3k\mathcal V_km_\text{P}^2}{8\pi}e^{(3\omega-1)x} \right],
\end{array}
\end{eqnarray}
where $m_\text{P}=1/\sqrt{G}$ is the Planck mass and the new
lapse function $\tilde N$ is defined by
\begin{eqnarray}\label{sh7}
\tilde N := \frac{N}{e^{3x}}.
\end{eqnarray}
Moreover,
\begin{eqnarray}\label{sh6}
p:=-\frac{3\mathcal V_ke^{3x}}{4\pi GN}\dot x,~~~~~p_\epsilon:=\frac{\mathcal V_ke^{3x}}{N^\frac{1}{\omega}(1+\omega)^\frac{1}{\omega}}(\dot\epsilon+\theta\dot{\mathcal S})^\frac{1}{\omega}e^{-\frac{\mathcal S}{\omega}},
\end{eqnarray}
are the conjugate momenta of the scale factor $x$ and $\epsilon$, respectively.
Note that the model has constraints of the
second kind, i.e., $p_\theta=0$ and $p_{\mathcal S}=\theta p_\epsilon$.

It is easy to show that employing the additional
canonical transformations\cite{Jalal}

\begin{eqnarray}\label{sh8}
\begin{array}{cc}
p_T:=-\frac{e^{\mathcal S}}{\mathcal V_k^\omega m_\text{P}}p_\epsilon^{1+\omega},~~~~~T:=m_\text{P}\mathcal V_k^\omega p_{\mathcal S}e^{-\mathcal S} p_\epsilon^{-(\omega+1)},\\
\\
\bar\epsilon:=\epsilon-(\omega-1)\frac{p_{\mathcal S}}{p_\epsilon},~~~~~\bar{p}_\epsilon:=p_\epsilon,
\end{array}
\end{eqnarray}
it simplifies the ADM Hamiltonian to
\begin{eqnarray}\label{sh9}
H_\text{ADM}=\tilde N\left[ -\frac{2\pi}{m_\text{P}^2\mathcal V_k}e^{3(\omega-1)x}p^2-m_\text{P}p_T-\frac{3k\mathcal V_km_\text{P}^2}{8\pi}e^{(3\omega-1)x} \right].
\end{eqnarray}
In equation \eqref{sh9}, it is seen that the momentum $p_T$ is the only remaining canonical variable (associated with the perfect fluid), which appears linearly. We should also note that the new variable $T$ is dimensionless. The super-Hamiltonian constraint of the model is given by
\begin{eqnarray}\label{sh10}
\mathcal H:=-\frac{2\pi}{m_\text{P}^2\mathcal V_k}e^{3(\omega-1)x}p^2-m_\text{P}p_T-\frac{3k\mathcal V_km_\text{P}^2}{8\pi}e^{(3\omega-1)x}=0.
\end{eqnarray}
We can then obtain the Wheeler--DeWitt equation by imposing the standard quantization conditions on the canonical momenta:
\begin{eqnarray}\label{sh10a}
\begin{array}{cc}
\hat x=x,~~~~\hat p=-i\partial_x,\\
\hat T=T,~~~~\hat p_T=-i\partial_T.
\end{array}
\end{eqnarray}
Moreover, we assume that the super-Hamiltonian operator annihilate the wave function
\begin{eqnarray}\label{sh11}
\begin{array}{cc}
i\partial_T\Psi(x,T)=\\
\\-\frac{1}{2M}e^{3(\omega-1)x}\left[\partial^2_x-\frac{3}{2}(\omega-1)\partial_x \right]\Psi(x,T)+\frac{3k\mathcal V_km_\text{P}}{8\pi}e^{(3\omega-1)x}\Psi(x,T),
\end{array}
\end{eqnarray}
where $M$ is the dimensionless `mass' parameter and we have used the Laplace--Beltrami operator ordering.
 Requiring that the Hamiltonian operator $H$ in the right hand side of (\ref{sh11}) to be self-adjoint, the inner product of two arbitrary wave functions $\Phi$ and $\Psi$ must take the form
\begin{eqnarray}\label{sh12}
\langle\Phi|\Psi\rangle=\int_{-\infty}^\infty e^{-\frac{3}{2}(\omega-1)x}\Phi^*(x,T)\Psi(x,T)dx.
\end{eqnarray}

Redefining the wavefunction as
\begin{eqnarray}
\psi(x,T):=e^{\frac{3}{4}(\omega-1)}\Psi(x,T),
\end{eqnarray}
the Wheeler--DeWitt equation (\ref{sh11}) transforms to
\begin{eqnarray}\label{sh11a}
\begin{array}{cc}
i\partial_T\psi(x,T)=\\
\\-\frac{1}{2M}e^{3(\omega-1)x}\partial^2_x\psi(x,T)+\left[\frac{9\pi(\omega-1)^2}{8\mathcal V_km_\text{P}^3}e^{3(\omega-1)x}+\frac{3k\mathcal V_km_\text{P}}{8\pi}e^{(3\omega-1)x}\right]\Psi(x,T),
\end{array}
\end{eqnarray}
with inner product given by
\begin{eqnarray}\label{sh12aa}
\langle\phi|\psi\rangle=\int_{-\infty}^\infty \phi^*(x,T)\psi(x,T)dx.
\end{eqnarray}
Consequently, the Wheeler--DeWitt equation (\ref{sh11}) becomes in the form of the Schr\"odinger equation $i\partial_t\psi=H\psi$.

For simplicity and in order to obtain analytical solutions, from now on, let us proceed our consideration with only the spatially flat universe $(k=0)$ filled with the stiff matter described with $\omega=1$. Therefore, for such a simple case, the Wheeler--DeWitt equation (\ref{sh11}) reduces to the Schr\"odinger equation of a free particle in $1D$ mini-superspace:
\begin{eqnarray}\label{sh13}
i\partial_T\Psi(x,T)=-\frac{1}{2M}\partial^2_x\Psi(x,T),
\end{eqnarray}
where $\mathcal V_0$ denotes the 3-volume associated with $k=0$.
The solution of the above `free particle' Schr\"odinger equation could be represented by a superposition of momentum eigenfunctions, with coefficients given by the Fourier transform of the initial wavefunction
\begin{eqnarray}\label{sh14}
\Psi(x,T)=\frac{1}{\sqrt{2\pi}}\int_{-\infty}^\infty\Phi_0(p)e^{i(px-\omega t)}dp,
\end{eqnarray}
where $\omega:=\frac{ p^2}{2M}$ and $\Phi_0(p)$ is the Fourier transform of the wavefunction $\Psi(x,0)$. Let us consider the
 initial state to be a normalized wave function
 \begin{eqnarray}\label{sh14a}
 \Phi_0(p)=\left(\frac{a^2}{2\pi}\right)^\frac{1}{4}e^{-\frac{a^2}{4}(p-p_0)^2},
 \end{eqnarray}
where $a,p_0\in\mathbb R$. Inserting the initial wave function (\ref{sh14a}) into (\ref{sh14}) and performing the
Gaussian integration, we obtain
\begin{eqnarray}\label{sh14b}
\Psi(x,T)=\frac{1}{\sqrt{q}}\left(\frac{a^2}{8\pi}\right)^\frac{1}{4}e^{ip_0(x-\frac{p_0}{2M}T)}e^{-\frac{1}{4q}(x-\frac{p_0}{M}T)^2},
\end{eqnarray}
where $q:=\frac{a^2}{4}+i\frac{T}{2M}$. Therefore, the expectation value of the scale factor is given by
\begin{eqnarray}\label{sh14c}
\langle x\rangle=\frac{p_o}{M}T=\frac{4\pi p_0}{m_\text{P}^2\mathcal V_0}T.
\end{eqnarray}

In what follows, we would like to extend our herein model by applying the most generalized fractional QM framework presented in subsection \ref{ST-FQM}. Concretely, the `space-time' fractional counterpart of the Schr\"odinger--Wheeler--DeWitt equation (\ref{sh11a}) is given by
\begin{eqnarray}\label{sh15}
\begin{array}{cc}
i^\beta\partial^\beta_T\psi(x,T)=
\frac{2\pi}{\mathcal V_km_\text{P}^3}e^{3(\omega-1)x}(-\partial_x^2)^\frac{\alpha}{2}\psi(x,T)\\
\\+\left[\frac{9\pi(\omega-1)^2}{8\mathcal V_km_\text{P}^3}e^{3(\omega-1)x}+\frac{3k\mathcal V_km_\text{P}}{8\pi}e^{(3\omega-1)x}\right]\psi(x,T),~~~~1<\alpha\leq2,~~0<\beta\leq1.
\end{array}
\end{eqnarray}
We should note that the time coordinate $T$ as well as $x$ are dimensionless and consequently we have assumed the scaling coefficients in `space-time' fractional SE (\ref{ST-1}) are unit, i.e.,  $\hbar_\beta=D_{\alpha,\beta}=1$. Again, assuming our simple case with $k=0$ and $\omega=1$, equation \eqref{sh15} reduces to
\begin{eqnarray}\label{sh15a}
i^\beta\partial^\beta_T\Psi(x,T)=
\frac{1}{2M}(-\partial_x^2)^\frac{\alpha}{2}\Psi(x,T),~~~~1<\alpha\leq2,~~0<\beta\leq1.
\end{eqnarray}
Applying the Fourier transform to the wavefunction as
\begin{eqnarray}\label{sh16}
\Psi(x,T)=\frac{1}{\sqrt{2\pi}}\int_{-\infty}^\infty dp e^{ipx}\Phi(p,T),
\end{eqnarray}
we obtain
\begin{eqnarray}\label{sh17}
i^\beta\partial_T^\beta\Phi(p,T)=\frac{1}{2M}|p^2|^\frac{\alpha}{2}\Phi(p,T).
\end{eqnarray}
The solution of the above equation is
\begin{eqnarray}\label{sh18}
\Phi(p,T)=E_\beta\left( \frac{i^\beta}{2M}|p^2|^\frac{\alpha}{2}t^\beta\right)\Phi_0(p),
\end{eqnarray} \label{Mittag}
where $\Phi_0(p)=\Phi(p,T=0)$ and
\begin{eqnarray}
E_\beta(z):=\sum_{n=0}^\infty\frac{z^n}{\Gamma(\beta n+1)},
\end{eqnarray}
is the Mittag--Leffler function. Inserting (\ref{sh18}) into (\ref{sh16}) yields the wavefunction of the `space-time' fractional stiff matter flat universe:
\begin{eqnarray}\label{sh19}
\Psi(x,T)=\frac{1}{\sqrt{2\pi}}\int_{-\infty}^\infty dp e^{ipx}E_\beta\left( \frac{i^\beta}{2M}|p^2|^\frac{\alpha}{2}t^\beta\right)\Phi_0(p).
\end{eqnarray}
If we choose the following `weight' function \cite{L00}, $\Phi_0(p)$ as the fractional generalization of Gaussian `weight' (\ref{sh14a})
\begin{eqnarray}\label{sh20}
\Phi_0(p)=\left(\frac{a\nu}{2^\frac{\nu+1}{\nu}\Gamma(\frac{1}{\nu})} \right)^\frac{1}{2}e^{-\frac{1}{4}a^\nu|p-p_0|^\nu},~~~~~p_0>0,~\nu\leq\alpha,
\end{eqnarray}
then the wavefunction becomes
\begin{eqnarray}\label{sh21}
\Psi(x,T)=\frac{1}{\sqrt{2\pi}}\left(\frac{a\nu}{2^\frac{\nu+1}{\nu}\Gamma(\frac{1}{\nu})} \right)^\frac{1}{2}\int_{-\infty}^\infty dp e^{ipx}E_\beta\left( \frac{i^\beta}{2M}|p^2|^\frac{\alpha}{2}t^\beta\right) e^{-\frac{1}{4}a^\nu|p-p_0|^\nu}.
\end{eqnarray}
Hence, the density of the probability of finding the universe to `occupy' a scale factor $x$ is
\begin{eqnarray}\label{sh22}
\begin{array}{cc}
|\Psi(x,T)|^2=\displaystyle\frac{a\nu}{2^\frac{\nu+3}{\nu}\pi\Gamma(\frac{1}{\nu})} \int_{-\infty}^\infty dp_1dp_2\\
\\
e^{i(p_1-p_2)x}E_\beta\left( \frac{i^\beta}{2M}|p_1^2|^\frac{\alpha}{2}t^\beta\right)E_\beta\left( \frac{(-i)^\beta}{2M}|p_2^2|^\frac{\alpha}{2}t^\beta\right) e^{-\frac{1}{4}a^\nu(|p_1-p_0|^\nu+|p_2-p_0|^\nu)}.
\end{array}
\end{eqnarray}
Note that for $\beta\neq 1$ the total probability is not conserved and it increases with time. For
large values of $\beta$ the total probability increases much faster.
For that `space' fractional case, $\beta=1$, the expectation value of the scale factor $x$ is
\begin{eqnarray}\label{sh23}
\langle x\rangle=\frac{\alpha p_0^{\alpha-1}}{2M}t,
\end{eqnarray}
 which shows that the center of the L\'evy wave packet (\ref{sh21}) moves with the group velocity $\frac{\alpha p_0^{\alpha-1}}{2M}$.

\vspace{1cm}


\section{Discussion and Outlook}
\label{DandO}

\indent

The bibliography of FQM\cite{B}, within the three approaches herewith pointed (namely, space-, time-, space-time-fractional), is now
relevant and significant. Fractional calculus has indeed 
become a 
computational powerhouse in physics and many engineering applications\cite{Ortigueira-1}. Endeavouring  towards
 cosmology (including in a quantum mechanical context) has also recently emerged\cite{GC,JF, mdpi}.

In the review paper herein presented, we
follow the `ouverture' made in another
publication, with an
emphasis on the canonical description\cite{mdpi}.
After a summary of FQM, we presented a 
{\it new} solution for a FRW model,
within the scope of {\it time}-FQM for a Schr\"odinger-Wheeler-DeWitt
equation.

Our results are depicted, also mentioning the 
corresponding ones within 
the standard Wheeler-DeWitt-SE for quantum cosmology; 
it can be easily appraised how the fractional feature (in the time derivative,
specifically) brings significant  alterations into the quantum dynamics of the universe. This assertion follows 
results in other publications\cite{mdpi} and herein, 
hence the suggestive 'omen' in our proposed title, broadening 
(canonical) quantum cosmology with a fractional agitation, like a 'whirl' of change. In particular, it is instructive to regard equations (\ref{Mittag}), (\ref{sh14c}), 
(\ref{sh22}), (\ref{sh23}) and contrast the standard to  the fractional `whirl'
brought into concrete observables.

As far as subsequent work, in the herein footsteps and elsewhere\cite{mdpi},
the covariant d'Alembertian (i.e., for both   time and space derivatives in FQM, physically 
consistent within geometrodynamics covariance) remains an open issue. So far, we have only
attempted `restricted' explorations for either space or time fractional derivatives, in separate, never together (surely not covariantly, as stressed).  These
limitations notwithstanding, the results so far are interesting and point that fractional calculus (even for simple case studies) can indeed open up
new routes towards more realistic and so far unchallenged situations
in quantum cosmology\cite{C2,VargasMoniz:2020hve,moniz2010quantum-a,moniz2010quantum-b}. The outlook, for (canonical) FQC is therefore warmly exciting and promising. The authors sincerely hope and trust that FQC will be a decisive route for exploration within
 the next decades.

\vspace{1cm}

{\bf Acknowledgments} The authors are grateful to J. Fabris and G. Calcani for letting us know of their earlier work. Likewise, we are most thankful to M. Ortigueira for sharing details about
his work and pointing to a rich collection of  references within
fractional calculus.  In addition, PVM and SMMR acknowledge the FCT grants UID-B-MAT/00212/2020 and UID-P-MAT/00212/2020 at CMA-UBI.  This article is based upon work conducted within  the  Action CA18108--Quantum gravity phenomenology in the multi-messenger approach--supported by the COST (European Cooperation in Science and Technology).

\end{document}